# Ultrathin All-Angle Hyperbolic Metasurface Retroreflectors Based on Directed Routing of Canalized Plasmonics

Li-Zheng Yin, Jin Zhao, Ming-Zhe Chong, Feng-Yuan Han, and Pu-Kun Liu*

State Key Laboratory of Advanced Optical Communication Systems and Networks,

School of Electronics, Peking University, Beijing, 100871, China

*Corresponding author: pkliu@pku.edu.cn

## Abstract

Retroreflectors that can accurately redirect the incident waves in free space back along their original channels provide unprecedented opportunities for light manipulation. However, to the best of our knowledge, they are suffering from either bulky size, narrow angular bandwidths, or time-consuming post-processing, which essentially limits their further applications. Here, a scheme for designing ultrathin all-angle real-time retroreflectors based on hyperbolic plasmonic metasurfaces is proposed and experimentally demonstrated. The physical mechanism underlying the scheme is the high-efficiency all-angle transition between the traveling waves in free space and the canalized spoof surface plasmon on the hyperbolic plasmonic metasurfaces. In this case, the strong confinement characteristic benefited from the enhanced light-matter interaction enables us to route and retroreflect the canalized spoof surface plasmon with extremely thin structures. As proof of the scheme, a retroreflector prototype with a thickness approximately equal to the central wavelength is designed and fabricated. Further experimental investigation obtains a half-power field of view up to 53° and a maximum efficiency of 83.2%. This scheme can find promising applications in target detection, remote sensing, and diverse on-chip light control devices.

**KEYWORDS:** all-angle retroreflection, spoof surface plasmon, hyperbolic plasmonic metasurfaces, canalization regime

## Introduction

Retroreflectors are a kind of novel optical or microwave mirrors that can efficiently reflect arbitrary incident waves back along their original path, in which you can see yourself at arbitrary angles. Their excellent capability in beam steering makes them ubiquitous in diverse industrial areas including free-space communication,[1-3] dynamic tags,[4] remote sensing,[5] and target detection.[6, 7] In the optical regime, the pyramid prism that performs two orthogonal specular reflections for each incident wave are the most classical structures to realize high-efficiency retroreflection.[3, 8-11] Moreover, this general design principle can also be scaled to the microwave and terahertz regime by replacing the mirrors with perfect electric conductors (PECs). However, as can be expected, their microwave and terahertz counterparts will be quite bulky due to the large electrical size and thus not natively compatible for integration and miniaturization. In this case, diverse schemes and structures are proposed and demonstrated to miniaturize the retroreflectors, in which metasurfaces stand out for their intrinsic ultrathin thickness.

Recently, metagratings, a new class of metasurfaces that rely on unique and periodically repeating unit cells to channel the incident power into the desired harmonics, emerge as new tools for light manipulation for their advantages of low mutual coupling, high efficiency, and easy processing.[12-16] In 2017, V.S. Asadchy *et.al.* propose and demonstrate a method of designing three- and five-channel retroreflectors by modulating the ultrathin metagratings with the specified surface impedance distribution.[17] Simplified structures on the basis of the principle of pattern multiplication are also demonstrated subsequently.[18, 19] To further increase the efficiency of the retroreflector, dielectric metagratings based on the generalized Kerker effect induced by interference between Mie-type resonances are also proposed.[20-25] These works can achieve the functionalities of high-efficiency retroreflection by elaborately designing the unit cells of the metagratings with specific scattering patterns. However, due to the inherent discrete properties of the scattering harmonics, retroreflectors designed by the periodic metagratings are theoretically confined to finite (< 3) incident directions and extremely narrow angular bandwidth. It means a sharp drop of retroreflection efficiency when the incident waves have a slight derivation with the specified incident direction. To overcome this shortcoming, in 2017, Amir Arbabi *et.al.* provide a new optical design framework to achieve wide-angle retroreflection with two cascaded metasurfaces and obtain a normal incidence efficiency of 78% and a large half-power field of view of 50°.[26] Similar structure is also designed and demonstrated at the telecommunication wavelength of 1550 nm.[27] However, to achieve the reported good performance, the distance between the two cascaded metasurfaces should be up to 590 $\lambda_0$. To simultaneously realize wide-angle retroreflection and miniaturization, in 2018, Libin Yan *et.al.* employ a single Pancharatnam–Berry metasurfaces to design a high-efficiency retroreflector that adaptively reflects the incident waves by mechanically altering the geometry of the unit cells.[28] This novel adaptive scheme requires post-processing and mechanical control and is thus time-consuming, limiting its further applications in real-time scenarios.

In this work, we propose a scheme to design ultrathin and all-angle retroreflectors based on hyperbolic plasmonic metasurfaces (HPMs). The premise of the scheme is to efficiently convert the incident waves with arbitrary incidence angles into the canalized spoof surface plasmon (SSP) on the HPMs, the process of which is reversible. Considering the strong confinement characteristic (field decays exponentially perpendicular to the propagation direction) benefited from the enhanced light-matter interaction, the canalized SSP thus can be routed and retroreflected with ultrathin planar structure. In this way, the retroreflection of the arbitrary incident waves in free space is converted to the retroreflection of the canalized SSP on the HPMs, which effectively reduces the size of the retroreflectors. As proof of the scheme, we design and fabricate a retroreflector that consists of an upper periodic metal cylinder array and a lower SSP retroreflector. The former aims at converting the incident waves into the canalized SSP and vice versa, while the latter composed of the normal and deformed HPMs is employed to route and retroreflect the canalized SSP. The designed retroreflector working at $f$ = 11.2 GHz has a thickness of 29.1 mm, which approximately equals its central operating wavelength $\lambda_0$ = 26.7 mm. Experimental investigation obtains a half-power field of view up to 53° and a maximum efficiency of 83.2%. The efficiency and angular bandwidth can be further increased by optimizing the structure parameters of the retroreflector, especially the metal cylinder array. This scheme can find promising applications in target detection, remote sensing, and diverse on-chip light control devices.

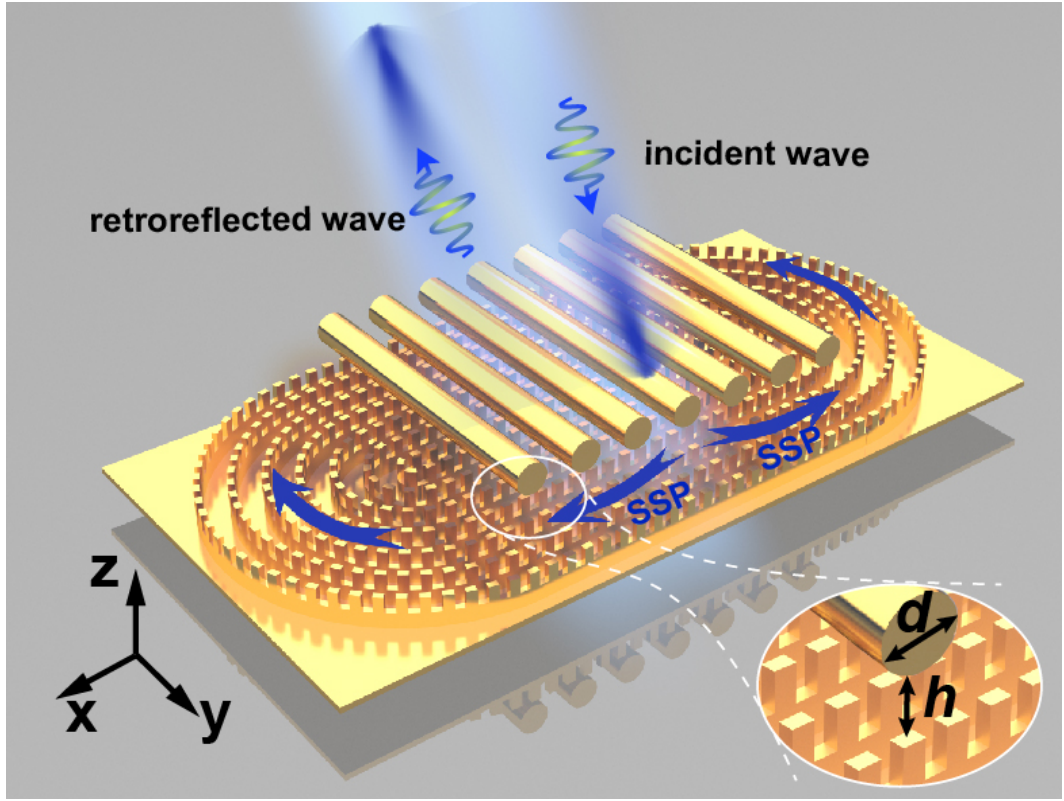

Figure 1. Schematic of the proposed retroreflector that can reflect the incident wave back along its original direction. The diameter and period of the upper cylinder are $d$ and $p$, and the distance between the metal cylinder array the HPMs is $h$.

## High-efficiency all-angle SSP coupling and decoupling

As shown in Fig. 1, the proof-of-scheme retroreflector consists of the lower normal and deformed HPMs, and the upper evenly arranged metal cylinders. Arbitrary electromagnetic waves with wave vectors laying in $yz$ plane will be reflected towards their original directions when it impinges on the retroreflector. The metal cylinder array above is utilized to modulate the incident waves into the canalized SSP with the Floquet wave vector $k_F$ in the $x$ directions and simultaneously demodulate the retroreflected SSP into the traveling wave towards the incident direction. The period of the array and the diameter of the cylinders are $p$ and $d$, and the distance between the metal cylinder array the HPMs is $h$.

As we mentioned in the Introduction, the first step of the proposed scheme is to efficiently convert the incident waves with arbitrary incidence angles into the SSP on the HPMs and vice versa, i.e., high-efficiency all-angle SSP coupling and decoupling. To realize this goal, the classical SSP waveguides such as the metal gratings and the conventional plasmonic metasurfaces cannot be the appropriate carriers of SSP due to their inherent 1D properties.[29-32] For these waveguides, there is only one or two specified directions for the incident wave that can be used to excite the SSP once the

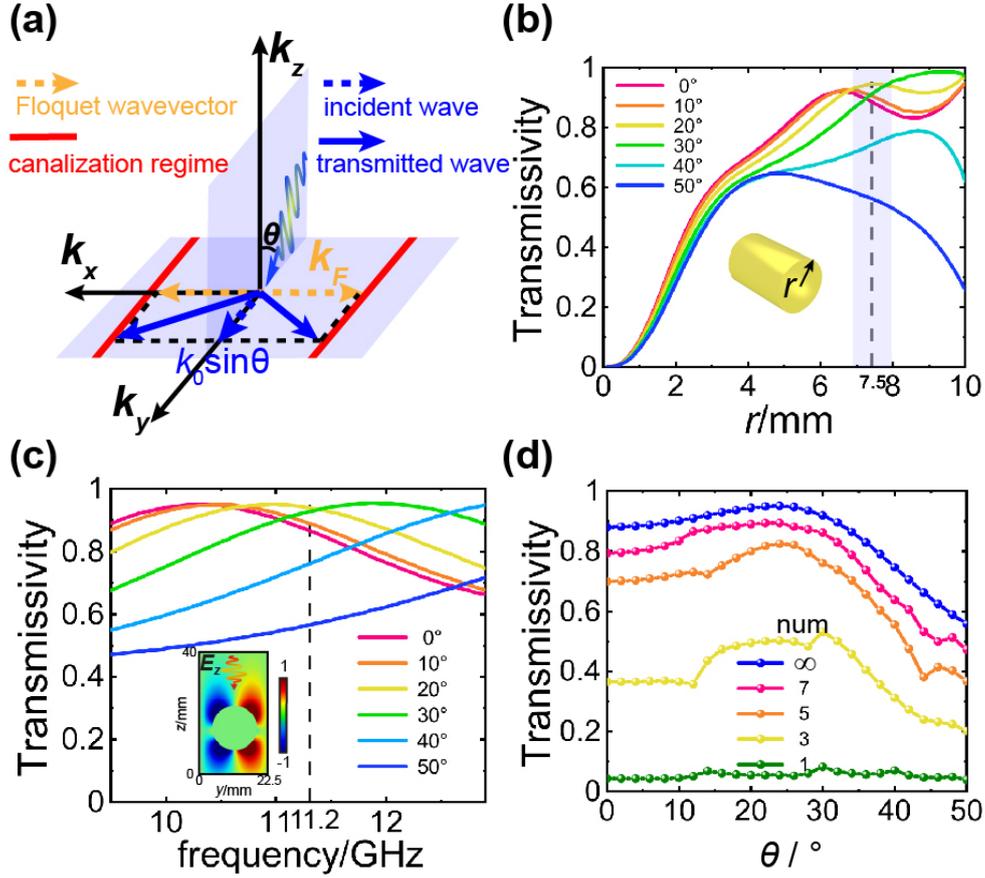

Figure 2 (a) Schematic of the tangential wave vectors of the all-angle coupling process. (b) The transmissivity of first-order harmonic of the desired first-order harmonic versus different incidence angles $\theta$ and radius $r$. (c) The transmissivity of the desired first-order harmonic versus different incidence angles $\theta$ and operating frequency $f$. The inset displays the electric field $E_z$ distribution under $\theta = 0°$ and $f = 11.2$ GHz. (d) The transmissivity of the desired first-order harmonic versus different incidence angles $\theta$ and the number the metal cylinders $num$.

structure is determined. It can also be understood by their classical $\omega$-$\beta$ dispersion relation where there is only one corresponding propagation constant $\beta$ for each specific angular frequency $\omega$. However, when we consider the 2D anisotropic HPMs that have independent dispersion relation in the orthogonal directions, the number of the propagation constant $\beta$ corresponding to each specific angular frequency $\omega$ is infinite.[33-37] This characteristic provides us the possibility to design the desired SSP waveguides to realize reciprocal all-angle SSP coupling and decoupling. According to the mode coupling theory, the condition of tangential wave vector match between the transmitted waves from the metal cylinder array and the Eigen SSP of the HPMs must be fulfilled to guarantee the high-efficiency coupling and decoupling. For the incident wave with incidence angle $\theta$ and wave vector laying in $yz$ plane, as shown in the blue dotted lines in Fig. 2(a), its tangential wave vector can be expressed as $\boldsymbol{k_{it}} = k_0 \sin\theta\, \boldsymbol{y}$, where $k_0$ represents the wave vector in free space. In this case, the tangential wave vector of the transmitted wave from the metal cylinder array can be written as $\boldsymbol{k_t} = \pm k_F\, \boldsymbol{x} + k_0 \sin\theta\, \boldsymbol{y}$, as shown in the blue solid lines in Fig. 2(a). The former part of $\boldsymbol{k_t}$ shown by the pink dotted lines is induced by the metal cylinder array while the latter part comes from the incident wave. Besides, they are also independent and orthogonal to each other. To realize the high-efficiency all-angle coupling and decoupling, it's obvious that the desired equal frequency contour (EFC) of the HPM should have the form $\boldsymbol{k_s} = \pm k_F\, \boldsymbol{x}$ ($k_F > 0$), as shown by the red solid lines in Fig. 2(a). In HPMs and hyperbolic metamaterials, the frequency ranges in which the corresponding EFCs are parallel to the axes are also called canalization regimes.[38-41]

  Firstly, we design the metal cylinder array so as to realize the high-efficiency transitions between the traveling waves in free space and the canalized SSP. According to the Floquet theory, the additional Floquet wave vector $k_F$ and the period $p$ of the metal cylinder array satisfy the relationship $k_F = 2n\pi / p$, where $n$ that can take arbitrary integers represents the order of the scattering harmonics.[38] In this work, we only consider the first-order transmitted harmonic ($n = 1$) which is expected to have a larger power proportion than other higher-order harmonics. Here, the $k_F$ which can take any value greater than $k_0$ is determined as $1.19\, k_0$. The period of the metal cylinder array can be thus calculated as $p = 2\pi / 1.19\, k_0 = 22.5$ mm. To evaluate the influence of the radius $r$ on the modulation effect, we calculate the transmissivity of the first-order harmonic versus $r$ under different incidence angles $\theta$, as shown in Fig. 2(b). In the simulation, the $x$-direction periodic boundary condition is added to one of the cylinders in the array to simulate an infinite array. A port radiating TM polarization (magnetic field is $x$-polarized) plane wave is used to illuminate the cylinder while another port that can absorb all the transmitted waves is set to calculate the scattering parameter $S_{21}$, as shown by the inset in Fig. 2(c). Spatial Fourier transform is then applied to the transmitted waves to calculate the proportion $\eta_m$ of the first-order harmonic. In this way, the transmissivity of the first-order harmonic can be finally obtained by the multiplication of $\eta_m$ and $|S_{21}|^2$. As shown in Fig. 2(b), the relationship between the transmissivity of the first-order harmonic and the radius $r$ has no consistent trend for different incident angles. In this work, we sum the six curves and take the radius corresponding to the maximum of the total transmissivity as the optimal value of $r$, the resultant value of which is 7.5 mm. It's noted that the optimal value may be different when we adapt other evaluation criteria. In Fig. 2(c), we also calculate and plot the transmissivity of first-order harmonic versus frequency under different incidence angles. The inset displays the electric field $E_z$ distribution under $\theta = 0°$ and $f = 11.2$ GHz. Similar to the data in Fig. 2(b), the curves in Fig. 2(c) also have no consistent trend. However, in a relatively wide range around the central frequency $f = 11.2$ GHz, the curve of transmissivity versus frequency is relatively flat for different incidence angles, implying a wide frequency bandwidth of the metal cylinder array. The theoretical analysis above is all based on the assumption of the ideal period. In the practical implementation, the finite number of the cylinders will also have a nonignorable influence on the transmissivity of first-order harmonic. Figure 2(d) plots the transmissivity of first-order harmonic versus incidence angles under a different number of cylinders. Obviously, the more the number of cylinders, the better the transmission effect. The blue curve is calculated under the ideal periodic boundary condition, which can be seen as the result of infinite cylinders. When the number increases to $num = 7$, its gap with the theoretical limit is small enough. Combining the analysis above, in this work, seven metal cylinders with diameter $d = 15$ mm and period $p = 22.5$ mm are chosen to be placed above the HPM to efficiently modulate the incident waves. It's worth noting that the coupling and decoupling of SSP are reciprocal and the results above also hold for the demodulation from the canalized SSP to outgoing wave towards the incident direction.

Next, we design the HPM to achieve the desired extremely-anisotropic dispersion characteristic, i.e., parallel EFC at the targeted frequency. The HPM in this work is constructed by the periodic arrangement of the unit cell illustrated in Fig. 3(a) along the $x$ and $y$ directions. The specific structure parameters of the unit cell are: $b_y = 1.5$ mm, $b_x = 3$ mm, $h_1 = 3.75$ mm, $h_2 = 5.25$ mm, $p_x = 4.5$ mm, and $p_y = 7.16$ mm. Considering only the goal of realizing canalization regime at the targeted frequency $f = 11.2$ GHz, the structure parameters of the unit cell are not unique and quite difficult to directly obtain. For the period $p_x$ of the unit cell in the $x$ direction, it can theoretically take any value smaller than the period $p$ of the upper metal cylinder array. However, a smaller $p_x$ that corresponds to a smoother field distribution of SSP always contributes to a better coupling effect in the practical implementation. In this work, it's determined as $p_x = 4.5$ mm, and the phase difference of the unit cell in $x$ direction at the targeted frequency can thus be calculated as $\Delta\varphi = 1.19\, k_0 p_x = 0.4\pi$. The situation is different for $p_y$ considering the match between the normal and the deformed HPMs, which will be discussed in the next section. For the uniaxial HPM in this work, its complete EFC can be determined once we obtain the dispersion characteristics of the SSP in the orthogonal directions.[43,44] For the convenience of analysis, the HPM can be approximately regarded as 1D metal grating for the SSP propagating only along $x$ or $y$ direction. In this case, the effective height and duty ratio of the slots of the metal gratings in $x$ and $y$ directions are $h_1$, $h_1 + h_2$, and $1 - b_x / p_x$, $1 - b_y / p_y$, respectively. To realize the canalization regime (EFC parallel to $y$ axis), the SSP propagating along $x$ direction should have the desired propagation constant $k_F = 1.19\, k_0$ at $f = 11.2$ GHz, while the SSP propagating along $y$ direction should have strong confinement so that its plasmonic frequency is far smaller than the targeted frequency $f = 11.2$ GHz. To achieve this goal, the effective height and duty ratio of the slots should be large in $y$ direction while relatively small in $x$ direction.

In this case, the final optimal values are $b_y$ = 1.5 mm, $b_x$ = 3 mm, $h_1$ = 3.75 mm, and $h_2$ = 5.25 mm. In the simulation, to obtain the complete EFC of the designed HPM, periodic boundary conditions with Floquet wave vectors $k_x$ and $k_y$ are added on the unit cell along x and y directions. By sweeping the combination of ($k_x$, $k_y$) with high resolution from (0, 0) to ($\pi / p_x$, $\pi / p_y$) and calculating the Eigen frequency at each sampling point of ($k_x$, $k_y$), the theoretical EFC of the designed HPM is finally obtained, as shown in Fig. 3(e). The cutting planes at 3.4 GHz, 6.8 GHz, and 11.2 GHz represent three typical shapes in the topological transformation of the EFC. Figures 3(b), 3(c), and 3(d) are the corresponding electric field $E_z$ distribution above the HPMs under the excitation of a vertical electric dipole. Figures 3(f), 3(g), and 3(h) are their practical EFC calculated by applying spatial Fourier transform on their corresponding electric

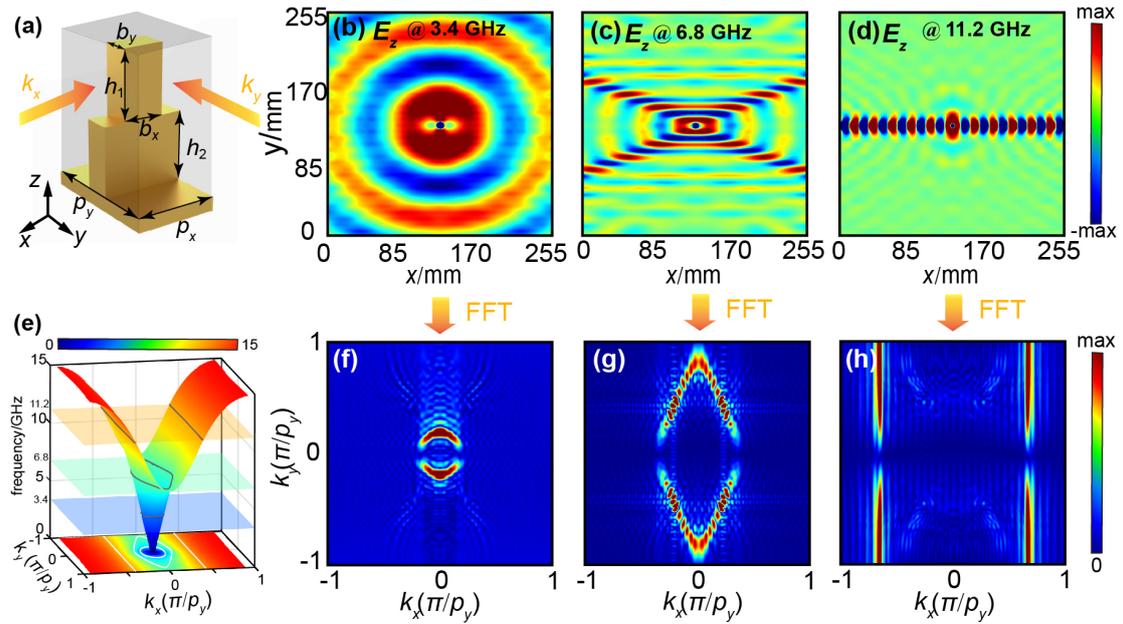

Figure 3 (a) Schematic of the unit cell of the HPMs. (b), (c), and (d) are the respective electric field $E_z$ distribution above the HPMs excited by electric dipoles at 3.4 GHz, 6.8 GHz, and 11.2 GHz. The cutting plane is 5 mm above the HPMs. (e) 3D and 2D color-filled contour maps of the EFC of the designed HPM. The cutting planes in the 3D map and the white lines in the 2D contour maps represent the theoretical EFC of the HPM at 3.4 GHz, 6.8 GHz, and 11.2 GHz. (f), (g), and (h) are the practical EFC of the HPM at 3.4 GHz, 6.8 GHz, and 11.2 GHz, respectively. They are calculated by applying spatial Fourier transform on the electric field $E_z$ distribution in (b), (c), and (d).

field $E_z$ distribution. As can be seen from Figs. 3(b), 3(e), and 3(f), when the operating frequency is small (< 4 GHz), the HPM can be regarded as an isotropic structure. This is because, compared with the large operating wavelength, the shape difference of the unit cell in the orthogonal directions is too small to be distinguished. The anisotropic dispersion characteristics gradually appear when the operating frequency increases to around 6.8 GHz, as can be seen from the rhombic EFC illustrated in Fig. 3(g). When the operating frequency is further increased, a topological transformation of EFC from rhombus to straight lines will occur. This canalization regime with the parallel EFC means that the corresponding SSP in this region (canalized SSP) with arbitrary transverse wave vector components has exactly the same group velocity. By combining the designed HPM with the metal cylinder array, the high-efficiency all-angle coupling and decoupling of SSP will finally be realized. It should be noted that the distance $h$ between them also affects the coupling and decoupling effects, and the optimal value is $h = 5.1$ mm.

## High-efficiency routing and retroreflection of canalized SSP

Having demonstrated the high-efficiency all-angle SSP coupling and decoupling, next, we employ the deformed HPMs to realize high-efficiency routing and retroreflection of the canalized SSP. As shown in Fig. 4(a), the proposed SSP retroreflector is constructed by a bent semicircular HPM. It is designed to achieve directed routing of the canalized SSP which is assumed to propagate along the grooves. Before giving the detailed structure parameters, we first analyze its principle of retroreflection. Figure 4(b) shows the simplified geometrical model of the SSP retroreflector in Fig. 4(a) that illustrates the process of retroflection of the canalized SSP. In Fig. 4(b), the semicircle on the left represents the SSP retroreflector while the other part represents the designed HPM in the last section. Considering the anisotropic dispersion characteristics of the HPM, most of the canalized SSP are extraordinary waves whose directions of phase velocity and group velocity do not overlap.[45] Here, we use the crossed gray arrows to represent the incident canalized SSP, where the group velocity is denoted by $S_i$ and the phase velocity is denoted by $k_i$, as shown in Fig. 4(b) The extraordinary canalized SSP incident on the two marked channels with radius $R_n$ and $R_m$ have different local initial phases, denoted by $\exp(-j\varphi_n)$ and $\exp(-j\varphi_m)$, respectively. The corresponding retroreflected waves on the exit side can be expressed as $T_n \exp(-j\varphi_n - j\Delta\varphi_n)$ and $T_m \exp(-j\varphi_m - j\Delta\varphi_m)$, respectively, where $T_i$, $(i = m, n)$ represents the transmission coefficients, and $\Delta\varphi_i$, $(i = m, n)$ represents the transmission phases. Assuming that the propagation constant of the canalized SSP on the HPM remains unchanged when it propagates to the SSP retroreflector, the transmission phase can be approximately estimated as $\Delta\varphi_i = \pi k_F R_i$, $(i = m, n....)$ Note that, retroreflection means the complete flip of the directions of both the phase velocity and group velocity, which is different from the specular reflection induced by the perfect electric conductor where only the vertical components will reverse. In Fig. 4(b), the condition $\Delta\varphi_n = \Delta\varphi_m$ should be fulfilled to guarantee an ideal retroreflection effect, which means the transmission phases of all the channels remain

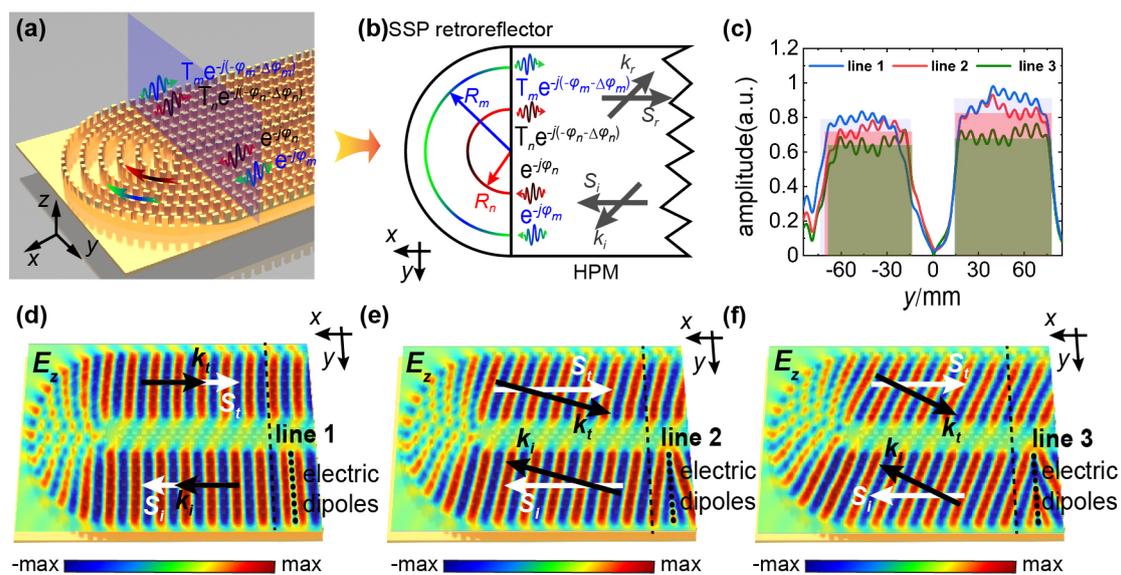

Figure 4 (a) Schematic of the deformed HPM which can retroreflect the canalized SSP with high efficiency. (b) Simplified geometrical model that illustrates the process of retroflection of the canalized SSP on the deformed HPM. (c) Amplitude distribution of electric field $E_z$ on lines 1, 2, and 3 in (d), (e), and (f). (d), (e), and (f) are the respective schematics of the retroreflection of the SSP with transverse wave vectors $k_y = 0$, $-0.26 k_0$, and $-0.5 k_0$, respectively. The electric field $E_z$ distribution excited by an electric dipole array with different phase gradient is used to simulate the field distribution on the HPM under the illumination of the incident waves with incidence angles $\theta = 0°$, $15°$, and $30°$, respectively.

the same. Considering the distance between the adjacent channels is $p_y$, the condition can be further simplified as $\pi k_F p_y = 2t\pi$, where $t$ represents an arbitrary positive integer. In this case, the smallest $p_y$ that meets this condition is $p_y = 2 / k_F = 7.16$ mm. This retroreflection process analyzed above is reversible and also holds for the canalized SPP impinging from the original exit side. To visualize and evaluate the practical effect of the SSP retroreflector, an electric dipole array densely arranged along $y$ direction with specific phase gradient $k_y$ is placed above the HPM to excite the canalized SSP. Figures 4(d), 4(e), and 4(f) are the respective electric field $E_z$ distribution above the HPM, visually showing the retroreflection of the canalized SSP with transverse wave vectors $k_y = 0$, $-0.26 k_0$, and $-0.5 k_0$. The black and white arrows mark the directions of phase velocity and group velocity of the canalized SSP, respectively, where subscript $i$ and $t$ represent the incident and retroreflected waves. The visual electric field distributions are in agreement with theory, confirming the ideal effect of the retroreflection. Besides, the unique characteristic of the canalized SSP can also be validated by the consistent group directions under different transverse wave vectors, as shown in Figs. 4(d), 4(e), and 4(f). Figure 4(c) plots electric field $E_z$ distribution on lines 1, 2, and 3 in Figs. 4(d), 4(e), and 4(f). Comparing the amplitude distribution on the incident and retroreflected sides, the total attenuation resulting from the transmission loss of the HPM and the radiating loss of the SPP retroreflector is overall acceptable. The accurate phase characteristics and the small retroreflection loss indicate the perfect performance of the SSP retroreflector.

## Experimental verification

To further demonstrate the effectiveness of the proposed scheme, a retroreflector prototype was fabricated by means of machining. As shown in Fig 5(a), the fabricated retroreflector is made of aluminum alloy 6061 with conductivity $\sigma = 2.77 \times 10^6$ S/m. Figures 5(c) and 5(d) show the partial zoom view of the upper aluminum cylinder array and the lower HPMs, respectively. In order to support the metal cylinder array, two holders with holes punched on them are fixed on both sides of the lower HPMs. For the lower HPMs, it has a size of 487 mm × 286.17 mm × 12 mm and there are 18 closed grooved paths machined on it. The size of the upper aluminum cylinder array is 195 mm × 269 mm × 15 mm. For the experimental configuration, the retroreflector is fixed vertically on the rotating stage while two horn antennas connected with the vector network analyzer Agilent N5245A are set in front of the retroreflector to emit and receive the signals, respectively, as shown in Fig. 5(b). Besides, the pyramidal and planar absorbers are also placed around and on the retroreflector to avoid the undesired

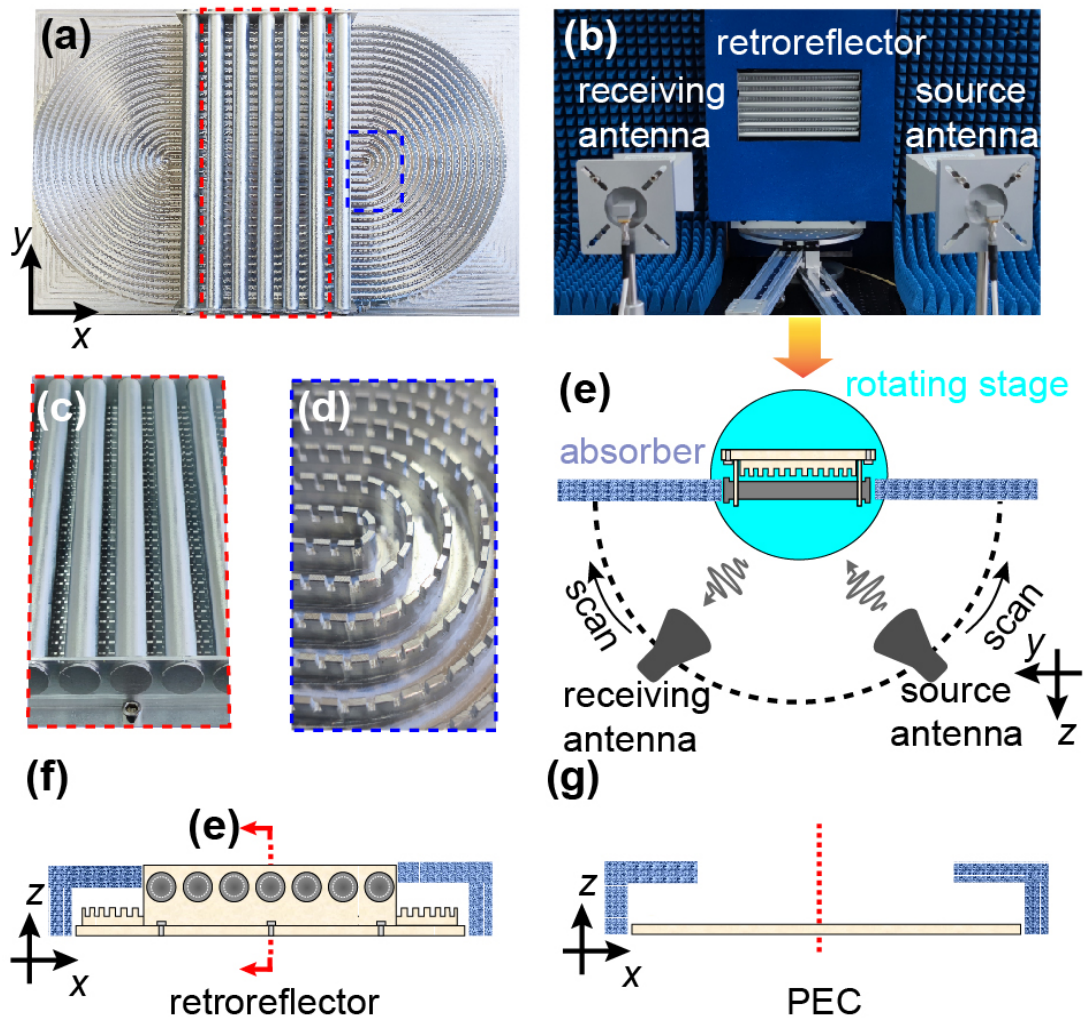

Figure 5 (a) Picture of the fabricated retroreflector prototype. (b) Picture of the measurement configuration, where the retroreflector under test is wrapped by the absorber except the metal cylinder array. (c, d) The partial zoom view of the retroreflector prototype. (e) Schematic of the vertical view of the measurement configuration. (f, g) Schematics of the side view of the retroreflector and PEC, respectively.

scattering. A rectangular hole with the size 200 mm × 260 mm (7.5$\lambda$ × 9.7$\lambda$) is tailored on the absorber above the aluminum cylinder array to allow electromagnetic illumination. Figure 5(e) shows the schematic of the vertical view of the measurement configuration in Fig. 5(b). The semicircular dotted line represents the scan path the horn antennas go along to test the retroreflector. To obtain the retroreflection efficiency $\eta(f)$ of the retroreflector, first, the source antenna is used to radiate the electromagnetic waves and simultaneously receive the retroreflection signal $R(f)$. The complete angular data can be obtained by continuously scanning along the circular path illustrated in Fig. 5(e). It is noted that the intensity of retroreflection signal $R(f)$ is much smaller than that of the self-reflection of the horn antenna caused by the nonideal match with the cable. Therefore, it cannot be directly obtained from the scattering parameter $S_{11}(f)$ of the retroreflector. To eliminate the interference, we can set the scattering parameter $S_{11}'(f)$ of the source antenna without load as memory data, and subtract it from the $S_{11}(f)$ of the retroreflector. In this way, the retroreflection signal $R(f)$ is effectively obtained. Then, as shown in Figs. 5(e), 5(f), and 5(g), by replacing the retroreflector with a PEC, we can obtain the scattering parameter $S_{21}(f)$ of the PEC with the same cross-section size using the receiving antenna. Figures 5(f) and 5(g) show the schematics of the side view of the retroreflector and PEC, respectively. In this case, the retroreflection efficiency $\eta(f)$ of the retroreflector can be finally calculated by normalizing the $R(f)$ by $S_{21}(f)$, i.e., $\eta(f) = R(f) / S_{21}(f)$.

Figure 6(a) shows the experimental results of the retroreflector at $f = 11.2$ GHz under different incidence angle $\theta$. The red and blue curves represent the amplitude of the post-processed retroreflection signal $R(f)$ of retroreflector and the specular reflection signal $S_{21}(f)$ of the PEC measured by the receiving antenna with the same cross-section size, respectively. Their difference representing the retroreflection efficiency $\eta(\theta)$ of the fabricated retroreflector is plotted by the black curve. The green shadow marks the angular region where the retroreflection efficiency is over 50%. As can be concluded from the results, maximum efficiency of 83.2% and a half-power field of view up to 53° is experimentally obtained, demonstrating the good retroreflection performance of the fabricated retroreflector prototype. Figure 6(b) shows the retroreflection efficiency versus frequency under different incidence angles $\theta$, from which we can find that obvious retroreflection effects can be observed in the frequency range from 10.9 to 11.5 GHz. Besides, the full-width at half-maximum (FWHM) of the retroreflector is almost the same for different incidence angle $\theta$. Note that, there are some evident ripples in all the curves of efficiency versus frequency. They are caused by the multiple reflections between the source antenna and the retroreflector under test and can be weakened by increasing the distance between them. Table I summarizes the comparison of the performance of the fabricated prototype with other retroreflectors proposed in some representative references. To the best of our knowledge, the existing methods of designing retroreflectors suffer from either the bulky size, narrow angular bandwidth, or time-consuming post-processing. Here, we use the ultrathin structure with the thickness of $1.09\lambda$ to realize the real-time retroreflection with a half-power angle bandwidth up to 53º and a maximum retroreflection efficiency up to 83.2%, which breaks the barriers among the size, angular bandwidth, and post-processing time.

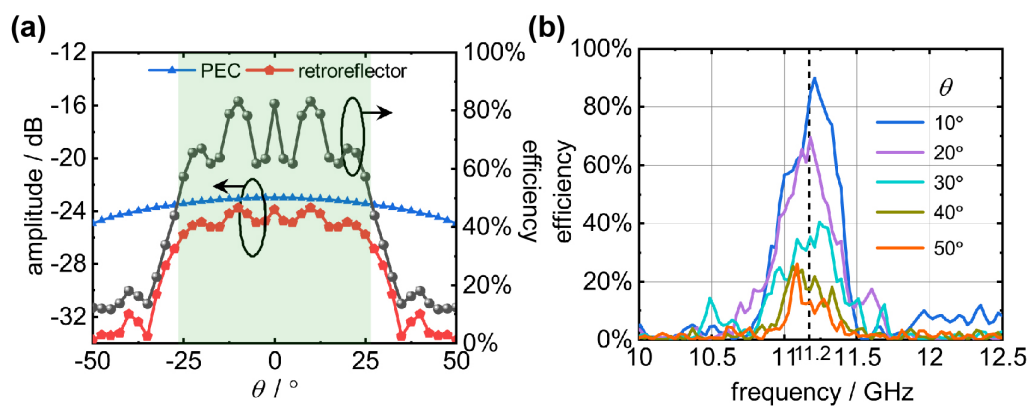

Figure 6 (a) Experimental results of the efficiency of the retroreflector under different incidence angle $\theta$. The red and blue curves represent the retroreflection coefficient (measured by the source antenna) of the retroreflector and the specular reflection coefficient of the PEC (measured by the receiving antenna) with the same size, respectively. Their difference representing the efficiency $\eta$ of the fabricated retroreflector is plotted by the black curve. (b) The experimental retroreflection efficiency of the fabricated retroreflector versus frequency under different incidence angles $\theta$, from which we can find that evident retroreflection effects can be observed in the frequency range from 10.9 to 11.5 GHz.

Table I. Comparison of The Performance of the Retroreflectors

| Retroreflector | Thickness | Central frequency | Half-power angle bandwidth | Maximum retroreflection efficiency | Post-processing |
|---|---|---|---|---|---|
| [17] | 0.04 $\lambda$ | 8 GHz | $\approx 0°$ | 90% | No |
| [18] | 0.125 $\lambda$ | 500 GHz | $\approx 0°$ | 99.6% | No |
| [20] | 0.33 $\lambda$ | Not given | $\approx 0°$ | 93% | No |
| [26] | 590 $\lambda$ | 560 nm | 50° | 78% | No |
| [27] | 276 $\lambda$ | 1550 nm | 50° | 96.8% | No |
| [28] | 0.2 $\lambda$ | 15 GHz | Not given | ~100% | Yes |
| This work | 1.09 $\lambda$ | 11.2 GHz | 53° | 83.2% | No |

## Conclusion and discussion

In this work, we propose a scheme to design ultrathin and all-angle retroreflectors based on HPMs. For practical implementation, the keys to the scheme are the high-efficiency all-angle coupling/decoupling and retroreflection of the SSP. As proof of the scheme, we design and fabricate a retroreflector working at $f$ = 11.2 GHz and it has a thickness of 29.1 mm, which approximately equals its central operating wavelength $\lambda_0$ = 26.7 mm. Experimental investigation obtains a half-power field of view up to 53° and a maximum efficiency of 83.2%. This scheme can find promising applications in target detection, remote sensing, and diverse on-chip wave control devices.

The efficiency and the angular bandwidth can be further increased by optimizing the structure parameters of the retroreflector. In this work, the main obstacle that limits the efficiency and the angular bandwidth is the relatively low transmissivity of the metal cylinder array. As can be seen from Fig. 2(b), the transmissivity drops drastically when the incidence angle exceeds 40°. The situation is worse considering the twice transmission in the process of the modulation and demodulation of wave vector $k_F$. It can be improved by changing the metal cylinder with phase-gradient metasurfaces. Besides, we can also employ high permittivity cylinders instead of metallic ones.

## Acknowledgments

This work was supported by National Key Research and Development Program under Grant No. 2019YFA0210203 and National Natural Science Foundation of China under Grant No. 61971013.

**Notes:** The authors declare no competing financial interest.